# Growth and optical properties of axial hybrid III-V/Si nanowires

Moïra Hocevar[1], George Immink[2], Marcel Verheijen[2,3], Nika Akopian[1], Val Zwiller[1], Leo Kouwenhoven[1] and Erik Bakkers[1,3, *]

[1]*Kavli Institute of Nanoscience, Delft University of Technology, 2628CJ Delft, the Netherlands*

[2]*Philips Innovation Services Eindhoven, High Tech Campus 11, 5656AE Eindhoven, the Netherlands*

[3]*Department of Applied Physics, Eindhoven University of Technology, 5600 MB Eindhoven, the Netherlands*

*\* Corresponding author: ebakkers@tue.nl*

**Hybrid silicon nanowires with an integrated light-emitting segment can significantly advance nanoelectronics and nanophotonics.[1-3] They would combine transport and optical characteristics in a nanoscale device,[4] which can operate in the fundamental single-electron[5] and single-photon regime.[6] III-V materials, such as direct bandgap gallium-arsenide, are excellent candidates for such optical segments.[7] However, interfacing them with silicon during crystal growth is a major challenge, because of the lattice mismatch, different expansion coefficients and the formation of anti-phase boundaries. Here, we demonstrate a silicon nanowire with an integrated gallium-arsenide segment. We precisely control the catalyst composition and surface chemistry to obtain dislocation-free interfaces. The integration of gallium-arsenide of high optical quality with silicon is enabled by short**

**gallium phosphide buffers. We anticipate that such hybrid silicon/III-V nanowires open practical routes for quantum information devices, where for instance electronic and photonic quantum bits are manipulated in a III-V segment[8] and stored in a silicon section.[9]**

**Introduction**

Silicon has a high thermal conductivity, is mechanically robust, and has a stable oxide. Its isotope $^{28}$Si has zero nuclear spin, which makes it especially attractive for quantum information devices, where a single electron spin is used as a quantum bit.[10] However, Si has an indirect bandgap, significantly limiting photonic applications. In contrast, most III-V's have a direct bandgap and therefore are perfectly suited for optoelectronic devices. Some, such as indium antimonide, also feature a high mobility for electrons and a very strong spin-orbit interaction, the ingredients for nanowire spin-orbit quantum bits (qubits).[11] Importantly, their electronic properties can be tuned by making more complex mixtures *e.g.* ternary or even quaternary compounds. For a wide variety of applications it is valuable to combine the best properties of Si and III-V semiconductors, but only if high quality heterojunctions can be realized. In bulk systems, threading dislocations are usually formed at the junction due to a mismatch in lattice constants and thermal expansion coefficients, even for a small lattice mismatch material combination like GaP/Si (0.3%).[12] These problems can be solved using bottom-up nanowires grown by the vapour-liquid-solid (VLS) mechanism, in which lattice strain is elastically relieved at the nearby surface.[13,14] Successful examples of high quality interfaces between heavily mismatched materials exist when both materials are from the same semiconductor group.[15-17] But defect-free interfaces between III-V and Si so far remain a challenge, even in nanowires.[18]

In this work we develop a method to integrate a gallium arsenide section in a silicon nanowire with atomically-sharp interfaces, and free of dislocations. Nanowires are grown via the VLS mechanism (see Methods Section) using gold particles. The gold particles catalyze the decomposition of the precursor molecules from the gas phase,[19] and then absorb the elements, forming a liquid alloy. When supersaturation is reached, atoms are expelled to form a crystal (either III-V or Si) below the liquid droplet. We aim at combining different classes of semiconductors, which are normally grown from their corresponding Au-based alloys, such as Au-Ga or Au-Si. To successfully grow III-V/Si heterostructured nanowires, the key is to control the surface chemistry and composition of the alloy.

We optimize the growth parameters and in particular for the switching between Si and the III-V semiconductor, which is a critical step. In order to switch from GaP to Si, a growth interrupt, during which phosphine is removed from the chamber, is essential, since phosphine hinders the absorption of Si in the catalyst particle. For switching from Si to GaP it is important to decrease the growth temperature, and increase the Ga pressure in the gas phase to expel Si from the catalyst alloy. Finally, an optically active GaAs segment is incorporated in the Si wire with good optical quality, as determined from photoluminescence measurements.

**Results**

**Optimisation of the GaP-Si junction.** First, to initiate uniform nanowire growth, GaP stems are grown at 540 °C on GaP(111) substrates from arrays of gold islands, and after a growth interrupt of several seconds, Si segments are grown on top. Fig. 1a shows a scanning electron microscopy (SEM) image of an array of 60 nm-diameter GaP-Si nanowires. We observe that most (95%) of the Si segments grow straight on top of GaP. Importantly, without the growth interrupt, realized by stopping the phosphine supply, we obtain kinked and defected Si segments on GaP[18] as shown in Fig 1b. The formation of kinked wires has previously been explained by island growth induced by the (solid-solid) hetero-interface energy.[18] In contrast, our results show that phosphine, or one of the decomposition products, kinetically hinders the absorption of Si in the Au particle (at the vapour-liquid interface), and reduces the supersaturation in the catalyst particle. After the growth interrupt the phosphine is desorbed, and Si atoms can efficiently be absorbed in the droplet. It has been observed before that phosphine affects the nucleation of Si wires,[20] and that at high enough Si supersaturation, proper layer-by-layer growth takes place. Since the solubility of group V elements in Au is very low, it is not expected that the solid-liquid interface energy is affected by the growth interrupt.

We statistically analyze kinking at the GaP-Si as a function of wire diameter and disilane pressure. For this study, arrays of Au islands are defined by electron beam lithography with varying particle size and interspacing. From topview SEM images (Figs 1c-f), straight and kinked wires are easily recognised as dots and short stripes. The kinking direction is not random, but always along three crystallographic [111] directions. The number of kinked wires increases with the wire diameter. This may either indicate that strain relaxation becomes inefficient or that the filling of the catalyst with silicon is slower for larger particles. The data, summarized in Fig 1g, also shows that the number of kinked wires has decreased by a factor of 10 by increasing the disilane pressure by a factor of 4, during the Si nucleation stage. This result shows that a high supersaturation in the catalyst particle is indeed beneficial for the growth of straight Si on GaP.

Fig. 2a shows a bright field transmission electron microscopy (TEM) image of a GaP-Si nanowire taken along a <110> zone axis. We observe two segments, GaP with stacking faults, and Si, which is free of planar defects (see Supplementary Fig. S1 for the image along a different <110> zone axis). From the contrast in the dark field image in Fig. 2b, we distinguish three different materials: the Au nanoparticle on top, the Si segment just below and the GaP segment at the bottom. This stacking is confirmed by energy-dispersive x-ray analysis (EDX) in Fig. 2c. From this EDX scan it is clear that the catalyst particle contains 35% Ga after cooling down, and the Si segment has thus grown from a Au/Ga alloy. We do not observe any effect of the Ga pressure, during the growth of GaP, on the Si growth. The interface between GaP and Si is atomically abrupt, as shown in Fig. 2d, confirming a layer-by-layer growth mechanism of Si on the flat GaP cross-section. We note that a thin Si shell (< 2 nm) forms around the GaP segment during Si growth (Fig. 2c), which is due to the relatively high temperature of the growth process.[21]

TEM measurements reveal that straight wires also can have stacking faults. Typically, for larger diameter wires, we find a set of two inclined twin boundaries in the Si segment (Fig 2e). These inclined stacking faults are always initiated at the solid-vapor interface of the GaP-Si junction. We present a statistical TEM analysis of these stacking faults in Fig. 2f, and show that for the optimum diameter of 60 nm about 80% of the wires is straight and defect-free. Important to note is that all kinked wires have these inclined stacking faults. A typical example is shown in Fig 2g. This indicates that the kinked segment has nucleated from the inclined twin plane. With a simple model (see supplementary Fig. S2), we intuitively explain the formation of the twin planes and the kink. An inclined facet can be formed at the triple-phase point, especially at low supersaturations.[22] Now, nucleation takes place either on the top facet or on the inclined facet. Depending on the contact angles of the catalyst, one of the two nucleation events will dominate. Nucleation on the inclined facet induces a twin and if the catalyst completely moves to the inclined facet a kink is formed.

**Optimisation of the Si-GaP junction.** In the next step we aim at fabricating the reversed junction; we grow a GaP segment on Si. The SEM image in Fig. 3a shows an array of uniform GaP-Si-GaP wires. We find an optimum yield (60%) of straight GaP segments grown on Si nanowires at a decreased growth temperature (500 °C) and increased trimethylgallium partial pressure with respect to the growth of the bottom GaP segment. By decreasing the growth temperature from 540°C to 500°C under $H_2$ we observe an increase in the yield of straight GaP segments on Si. We attribute the improved yield to the reduction of the Si concentration in the droplet. Moreover, an excess of Ga is needed in the Au particle to initiate GaP nucleation.[23] From the TEM images (see Figs. 3b,c) it is clear that the GaP segment grows epitaxially on Si. Moreover, the interface between the Si and the GaP segment is atomically abrupt. Under these growth conditions, GaP grows with the zincblende crystal structure and contains stacking faults perpendicular to the growth direction. We also observe a 45% increase in the diameter of the GaP segment compared to that of Si, which is partly explained by an increased particle volume (by the uptake of Ga), and by sidewall growth at these higher temperatures.[24] However, the most important contribution to the diameter increase comes from differences in contact angle between the Au/Ga alloy on a GaP and on a Si surface. This difference is due to the different surface energies at the solid-vapour, solid-liquid, and liquid-vapour interfaces for the Au/Ga-GaP and the Au/Ga-Si systems.[25]

**III-V/Si superlattice nanowires**. The possibility of stacking Si on GaP, and reverse, makes it possible to grow III-V/Si superlattice nanowires. Figs. 3d,e show examples of triple and quadruple GaP-Si heterostructures. The heterostructures are straight and show the typical modulation in diameter that is observed when switching from GaP to Si and reverse. An arbitrarily switching scheme is used here and we note that the segment lengths can be optimised on demand for electronic, photonic or phononic applications.

Such applications require materials with specific dimensions, which are controlled by the growth rate. Therefore the Si and GaP growth rates are studied as a function of the gold catalyst diameter and the position in the vertical direction along the nanowire (Fig. 3f). The growth rates of the upper segments are lower than those of the bottom segments, showing that precursor diffusion along the nanowire facets is rate limiting. The growth rate of the GaP wires is higher for smaller diameters, which is consistent with a constant collection area per wire.[26-27] In contrast to GaP, the Si growth rate increases with the catalyst diameter.[28-30] Two independent factors explain this behaviour: synergetic nanowire growth[19] and the Gibbs-Thomson effect.[29,31] We observe an increase (>15%) of the Si growth rate with decreasing wire-to-wire distance for a constant diameter, which we attribute to synergetic nanowire growth (see Supplementary Fig. S3). The synergetic effect is only relevant for Si interdistances below 1 µm. Since the results shown in Fig. 3f are obtained at a larger interdistance (2 µm), this effect can be ignored. A more important contribution comes from the Gibbs-Thomson effect known for increasing the silicon partial pressure in the catalyst upon decreasing the diameter. As a consequence, the incorporation of atoms from the gas phase into the catalyst is reduced.[28-29] We note that the silicon supersaturation inside the catalyst does not only depend on the catalyst radius, but also on the liquid-vapour surface energy. In our case, the silicon precursors do not alloy with pure Au, but with a Au/Ga (65/35) mixture (Fig. 2c). Au/Ga droplets have a much higher (> factor 3) surface energy,[25] shifting the critical diameter, below which no nanowire growth will occur, to higher values. We find a critical diameter of 12 nm (see Supplementary Fig. S4 and Supplementary Table S1). This is 4 times larger than the value found for Si[28] and Ge[31] nanowires, which is consistent with the higher liquid-vapour interface energy. Generally, it shows that the Gibbs-Thomson effect and the growth dynamics can be tailored by the catalyst chemical composition and surface energy.

**Si wires with an optically active GaAs segment.** We now focus on the incorporation of direct bandgap GaAs into Si nanowires using GaP barriers. The yield of straight Si-GaP-GaAs-GaP-Si (hybrid Si/GaAs) nanowires exceeds 50% (Fig. 4a). The short GaAs segment is visible in the middle of the dark-field image in Fig. 4b and measures 20 nm in length. The complex materials stack is confirmed by the EDX scan along the wire in Fig. 4c. Important to mention is that the GaAs segment is capped with thin GaP and Si shells with thicknesses of 3 nm and 6 nm, respectively, as calculated from the EDX line scans. The GaAs segment has crystallized in the zincblende crystal structure, containing twin planes and wurtzite sections[24] (see TEM images in supplementary S5).

We assess the optical quality of our hybrid Si/GaAs nanowires by time-resolved photoluminescence spectroscopy as presented in Fig. 5. The emission from a single GaAs section, with a length exceeding 1 µm, results in a bright and spectrally narrow peak at 1.495 eV (829.5 nm) with full width half maximum of 2.5 meV (see inset). The 25 meV redshift of the photoluminescence peak, as compared to bulk GaAs can be attributed either to carbon acceptor recombination[32] or to carrier confinement at the interfaces between zincblende and wurtzite crystal phases of GaAs.[33] We determine a lifetime of 7 ns, which is longer than usual exciton lifetimes in pure zincblende GaAs nanowires. This long lifetime originates from indirect recombination of electrons and holes, which are spatially separated at wurtzite/zincblende junctions in the GaAs segment.[34] These long lifetimes lead to an important conclusion - a high purity of the GaAs section. Defects or surface states in a nanowire that typically contribute to very fast nonradiative decay channels are negligible in our hybrid Si/GaAs nanowires.

**Discussion**

We have demonstrated epitaxial integration of III-V materials and Si in nanowires. We believe that our approach is generic and allows combining a broad range of semiconductors in an arbitrary sequence, opening new possibilities for device applications. For example, the detection of single photons in the infrared range could be enhanced by the introduction of a low band gap III-V semiconductor such as GaSb or InAs in the absorption region of Si avalanche nanowire photodiodes.[35] These nanowires have great promise in the field of quantum information processing, for instance spin-orbit quantum bits have recently been shown in III-V wires.[8] However, the electron spin coherence time in this system is relatively short because of interaction with the nuclear spins. A solution could be to integrate a $^{28}$Si segment in a III-V wire as a spin qubit memory element, combining fast qubit operation in III-V[8] and the long coherence time of the spin states in Si.[9]

**Methods**

**Patterns.** Regular patterns of gold nanoparticles are prepared by electron beam lithography (100 keV). We define particle diameters ranging from 10 nm to 250 nm. For each diameter, arrays of 25x25 holes are designed with pitches ranging from 200 nm to 5 μm. The GaP(111)B substrate is prepared prior to resist deposition by etching the surface in a solution of HCl:HNO$_3$:H$_2$O (3:2:2) at 50°C for two minutes. The resist, a commercial solution of polymethyl methacrylate (PMMA) with 950-k molecular weight, is spin-coated and baked at 175°C for 15 min. After writing the patterns, we develop the PMMA in a solution of Methyl isobutyl ketone:isopropanol (1:3). A thin layer of 6 nm of gold is evaporated in a vacuum chamber of a electron beam evaporator at a base pressure of 10$^{-8}$ Torr. We realize the lift-off at 70°C in a commercial solution of PSR3000 for 15 min. Finally, two baths of acetone and one of IPA are required to remove the remaining traces of resist.

**Nanowire growth**. The nanowires are grown in a horizontal Aixtron 200 MOVPE reactor with a total pressure of 25 mbar by the Vapour Liquid Solid (VLS) growth mechanism using Au particles, as described above. Before entering the reactor, the samples are chemically treated using a piranha solution. Prior to growth, we realize annealing at high temperature for decontamination purposes. The growth temperature for GaP-Si and GaP-Si-GaP nanowires is set between 480-610 °C. Above 540 °C, a thin Si shell is formed due to competition between VLS and surface growths. GaAs insertions are grown at temperatures in the range of 450-540 °C. The precursor flow rates are set to $2\times10^{-4}$ – $2\times10^{-3}$ mbar for trimethylgallium (TMG) and $2\times10^{-2}$ – $10\times10^{-2}$ mbar for phosphine ($PH_3$). The growth interrupt, which consists of removing the $PH_3$ flow prior to Si growth, is optimized at 3 seconds. While growing GaAs segments, the $AsH_3$ flow is $8\times10^{-3}$ mbar. We studied the influence of P, the $Si_2H_6$ partial pressure, at P, 2P and 4P with P = $6\times10^{-3}$ mbar.

**Micro-photoluminescence.** Micro-photoluminescence studies are performed at 4.2 K. We excite the nanowires with 780 nm picosecond pulsed lasers focused to a spot size of 1 μm using a microscope objective with numerical aperture 0.85. We use an excitation power of 200 nW. We collect the photoluminescence signal by the same objective and send it to a spectrometer, which disperses the photoluminescence onto a nitrogen-cooled silicon array detector or streak camera, enabling 30 μeV spectral and 20 ps temporal resolution, respectively.

**Acknowledgements**

The work and results reported in this publication were obtained with research funding from the European Community under the Seventh Framework Program for the Marie Curie Fellowships Intra-European-Fellowships (IEF) for career development "NanoSpid2" (Contract No. 253243) and from the Netherlands Organization for Scientific Research (NWO-VICI 700.10.441). We acknowledge S. Frolov for valuable discussions and critical reading of the manuscript.

**Author contributions**

M. H, M. V., L. K. and E. B were involved in the design of the experiments. L. K. and E. B. provided guidance throughout the project. M. H. and G. I. were responsible for nanowire growth. M. H. took the lead in the SEM analysis, and M. V. in the TEM analysis. M. H. and N. A were responsible for the optical measurements. All authors contributed to the writing and editing of the manuscript.

**Competing financial interests**

The authors declare no competing financial interests.

**Additional information**

Supplementary information accompanies this paper on www.nature.com/naturecommunications. Reprints and permissions information is available online at http://npg.nature.com/reprintsandpermissions. Correspondence and requests for materials should be addressed to E.B. (e.p.a.m.bakkers@tue.nl)

**Figure captions**

**Figure 1. GaP-Si nanowire arrays. a.** SEM image of an array of 60 nm diameter GaP-Si nanowires with GaP and Si segment lengths of 250 nm and 300 nm, respectively. Tilt angle=45°, scale bar = 1 µm. **b.** Cross section SEM image of an array of 60 nm diameter GaP-Si nanowires with GaP and Si segment lengths of 1µm and 500 nm, scale bar = 1 µm. **c-f.** SEM top-view images of GaP-Si nanowire arrays for different diameters and grown with a $Si_2H_6$ partial pressure of 2P=1.2x10$^{-2}$ mbar. The pitch is 500 nm for all images, scale bar = 3µm. The diameters are **c.** 40 nm **d.** 70 nm **e.** 100 nm and **f.** 120 nm. A dot corresponds to a straight wire. Clearly, the number of kinked wires increases with increasing diameter. **g.** Summary of the data for different $Si_2H_6$ partial pressures (P=0.6x10$^{-2}$ mbar) showing the fraction of straight wires. Each data point is obtained from a field of 625 wires.

**Figure 2. Structural properties of GaP-Si nanowires. a.** Bright-field TEM image of a single GaP-Si nanowire, scale bar = 50 nm. **b.** High angle annular dark field (HAADF) image of a single GaP-Si nanowire, scale bar = 100 nm. **c.** EDX line scan taken along the nanowire shown in (**b**). **d.** High-resolution scanning TEM - HAADF image of the interface between GaP and Si, showing the atomically abrupt interface, scale bar = 5 nm. **e.** Bright-field TEM of a straight but faulted GaP-Si nanowire. The twin is clearly visible, scale bar= 100nm. **f.** Plot of the number of straight versus faulted wires observed by TEM. **g.** Bright-field TEM of a kinked GP-Si nanowire with visible twins, scale bar = 200 nm.

**Figure 3. From GaP-Si-GaP multijunctions to superlattice nanowires. a.** SEM image of an array of 60 nm-diameter GaP-Si-GaP nanowires with GaP, Si and GaP segment lengths of 120 nm, 160 nm and 280 nm, respectively. Tilt angle = 80°, scale bar = 1 µm. **b.** TEM image of a single GaP-Si-GaP nanowire with a diameter of 28 nm, scale bar = 200 nm. **c.** High-resolution TEM picture of a Si-GaP transition, scale bar = 5 nm. Assembled SEM images of (**d**) triple GaP-Si heterostructures (Tilt angle=80°) with diameters of 33, 46 and 60 nm (left to right), and (**e**) quadruple GaP-Si heterostructures (Tilt angle=45°) with same diameters, scale bars = 500 nm. **f.** Growth rates of GaP and Si segments for nanowires with the various catalyst particle diameters and with respect to their position along the heterostructures in (**e**).

**Figure 4. GaAs segments in Si nanowires. a.** SEM picture of an array of GaP-Si-GaP-GaAs-GaP-Si (hybrid Si/GaAs) nanowires. Tilt angle=45°, scale bar = 1 µm. **b.** HAADF image of a 25 nm-diameter Si-GaP-GaAs-GaP-Si nanowire, scale bar = 200 nm. **c.** EDX scan taken along the nanowire growth axis in (**b**).

**Figure 5. Time-resolved photoluminescence spectroscopy of a single hybrid Si/GaAs nanowire measured at 4.2 K.** Color-coded photoluminescence intensity is plotted as a function of the emission wavelength and detection time

after excitation by a laser pulse impinging at time 0. The data at negative times represent the part of the photoluminescence decay that was excited by the previous laser pulse 13 ns earlier. The upper part represents the photoluminescence intensity integrated over the full time window of 2 ns. The right part represents the decay of the photoluminescence spectrum integrated between 825 and 835 nm. The lifetime was determined by extracting the slope of the extrapolated photoluminescence decay curve. Inset: photoluminescence of the same nanowire detected by a nitrogen-cooled silicon array detector.

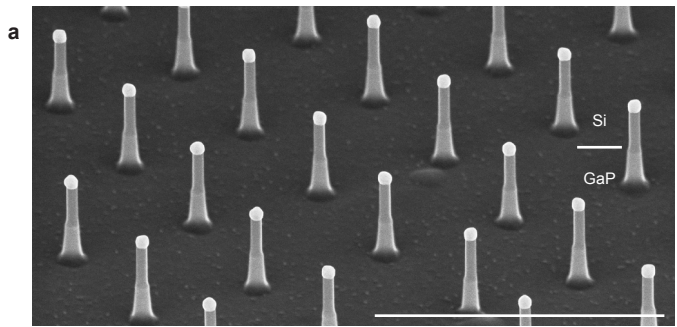
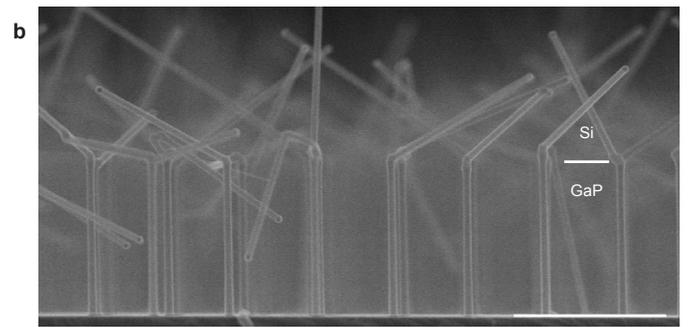
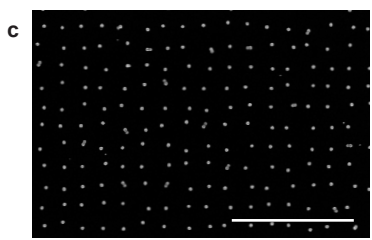
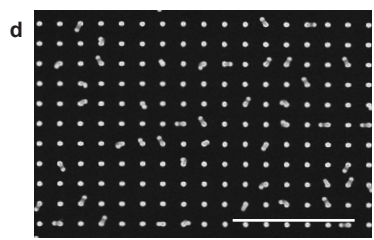
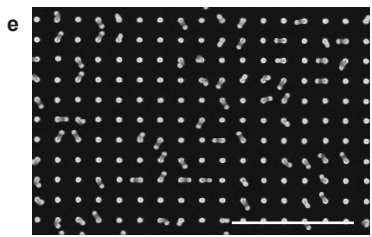
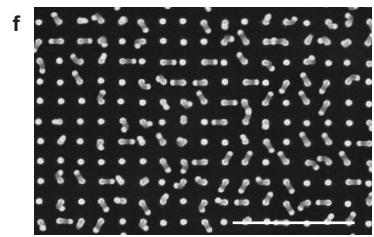
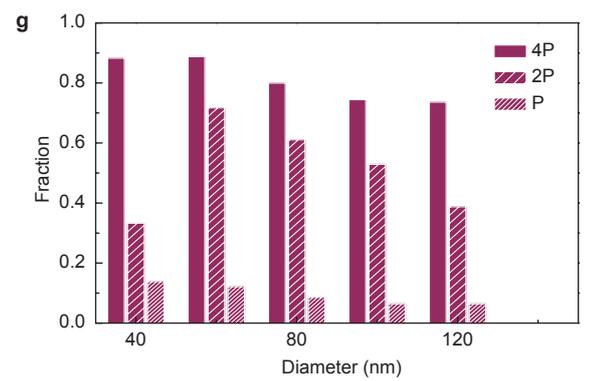

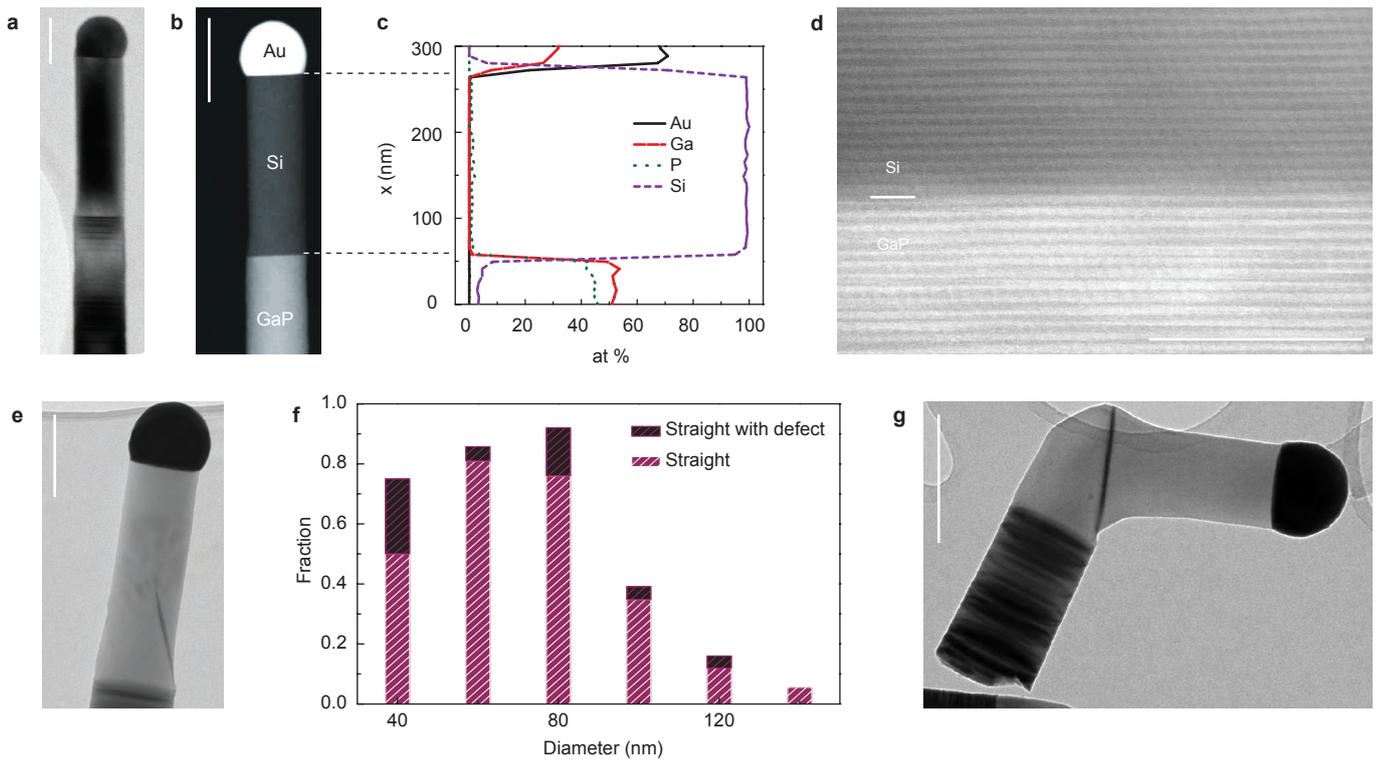

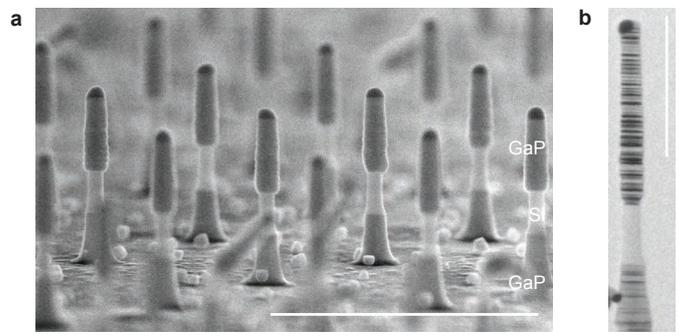

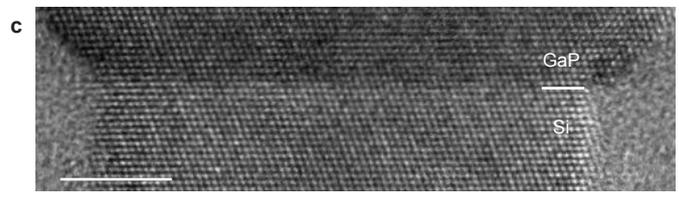

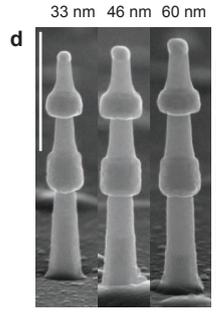

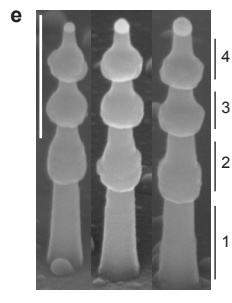

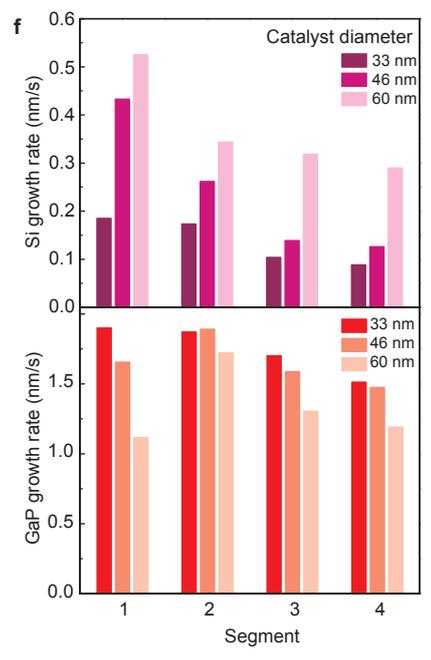

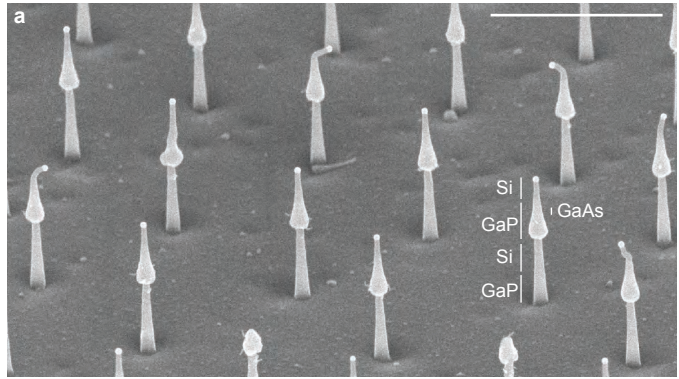
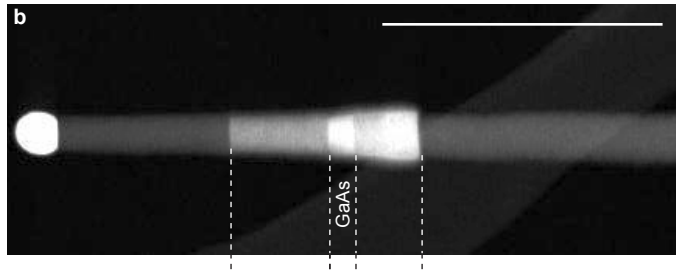
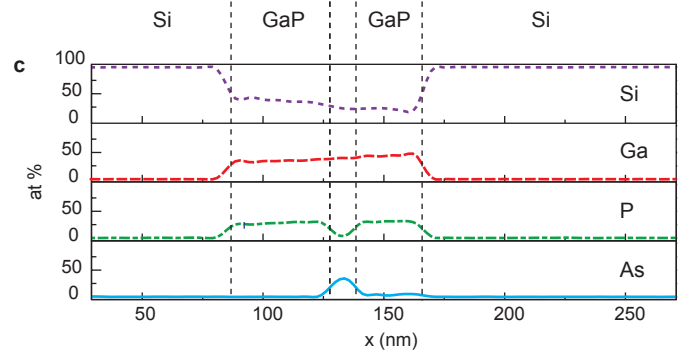

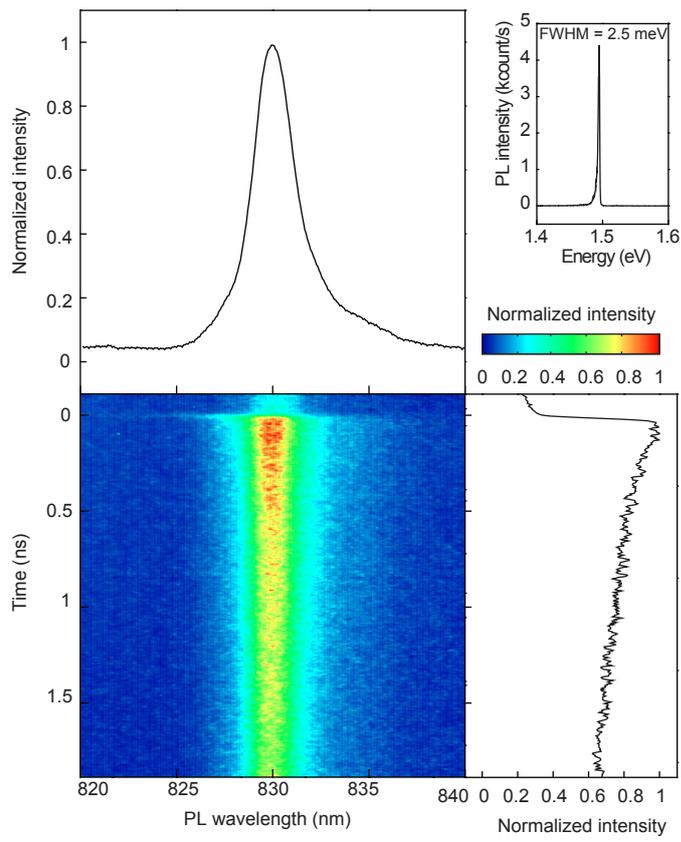

# Supplementary information

# Growth and optical properties of axial hybrid III-V/Si nanowires

Moïra Hocevar[1], George Immink[2], Marcel Verheijen[2,3], Nika Akopian[1], Val Zwiller[1], Leo Kouwenhoven[1] and Erik Bakkers[1,3, *]

*[1]Kavli Institute of Nanoscience, Delft University of Technology, 2628CJ Delft, the Netherlands*

*[2]Philips Innovation Services Eindhoven, High Tech Campus 11, 5656AE Eindhoven, the Netherlands*

*[3]Department of Applied Physics, Eindhoven University of Technology, 5600 MB Eindhoven, the Netherlands*

* Corresponding author: ebakkers@tue.nl

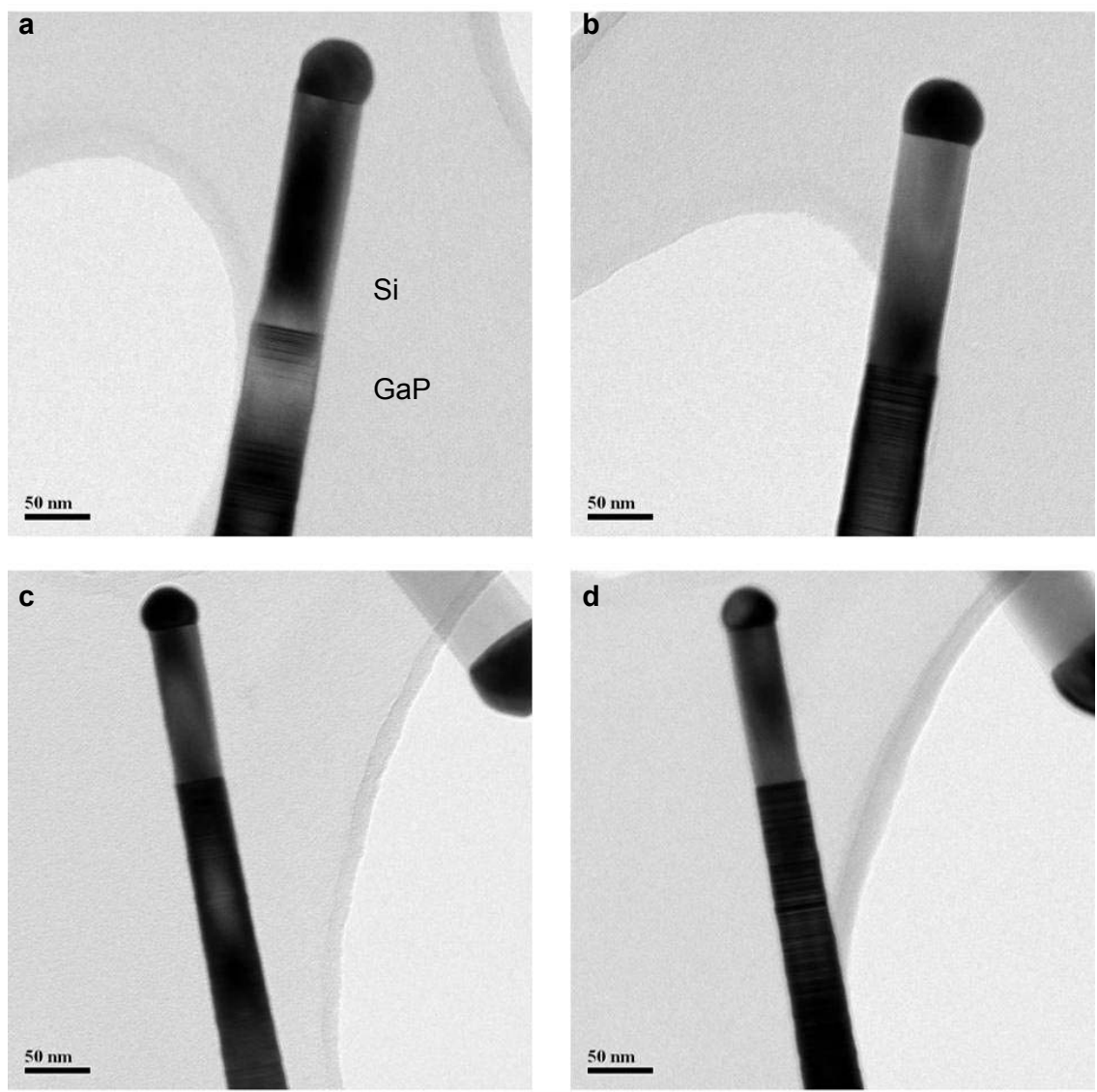

**Figure S1.** Bright-field TEM images of two different GaP-Si nanowires along the two different <110> zone axes, differing by 120° in rotation around the long axis. The diameters of the nanowires are **a-b**, 50 nm and **c-d**, 38 nm. Clearly, the Si nanowires are free of planar stacking faults.

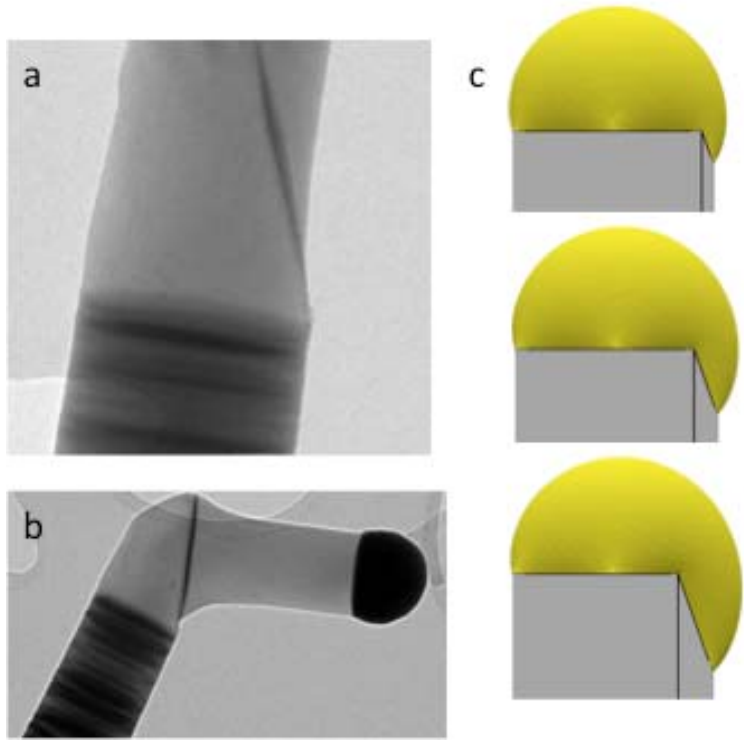

**Figure S2. a-b.** typical bright-field TEM images of two different GaP-Si nanowires. Both wires have a set of 2 inclined twin boundaries. The wire in **(a)** is straight while the wire in **(b)** is kinked**.** Important to note is that all wires, which kinked have these inclined stacking faults. The kinked segment has nucleated from the inclined twin plane. **c.** The wetting of the catalyst particle is calculated by using 'Surface Evolver' in case such an inclined twin plane is formed. The particle wets two facets. Now nucleation can take place either on the top facet or on the inclined facet and depending on the contact angles one of the two will dominate.

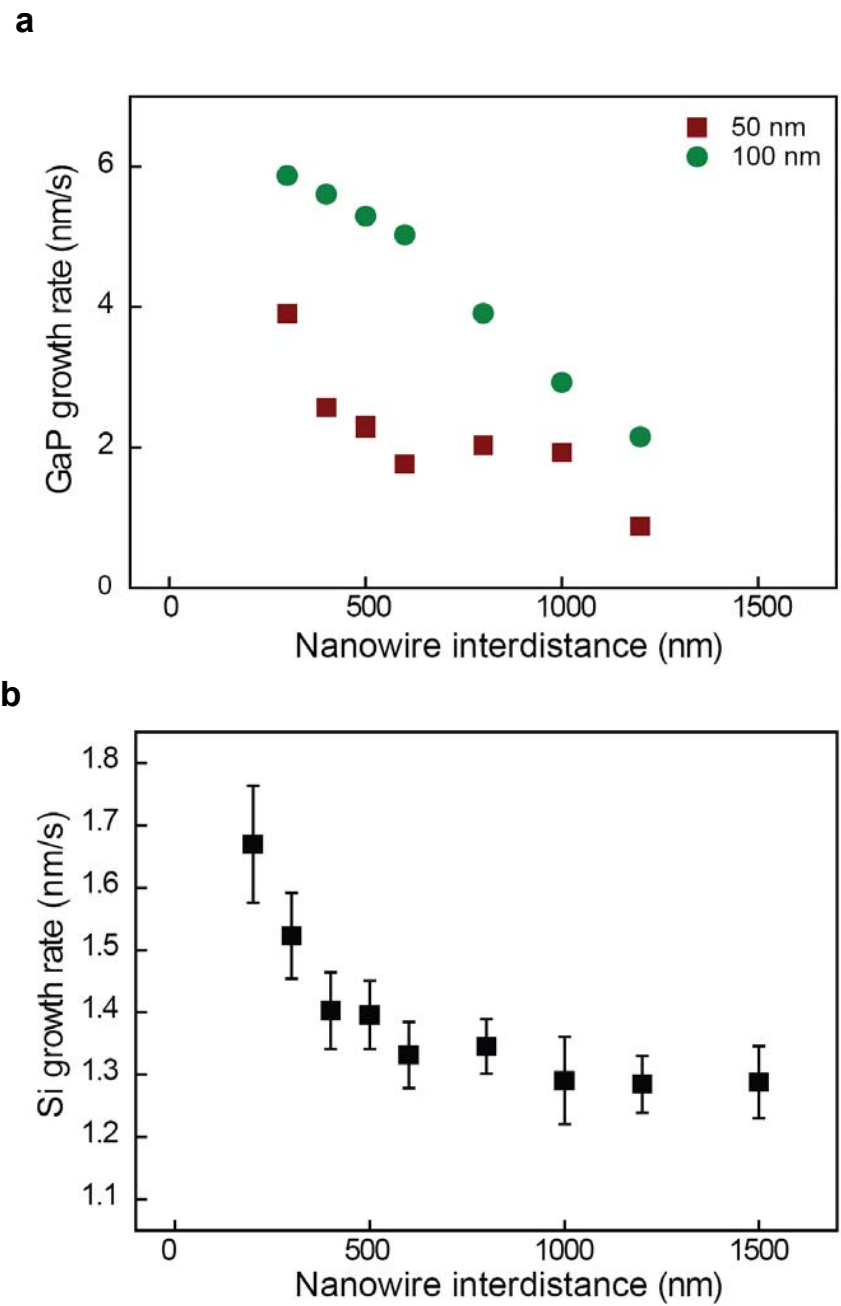

**Figure S3. a.** plot of the growth rate of **GaP** nanowires versus nanowire interdistance. It is clear from this plot that above 1 µm interdistance, the growth rate is constant, demonstrating independent growth between nanowires.[19] **b.** plot of the growth rate of **Si** nanowires versus interdistance between nanowires for a diameter of 100 nm. As shown on the figure, the maximal Si length variation (15%) occurs for 300 nm nanowire interdistance and is due to synergetic effects. At interdistances, above 1 µm, the synergetic effect is negligible.

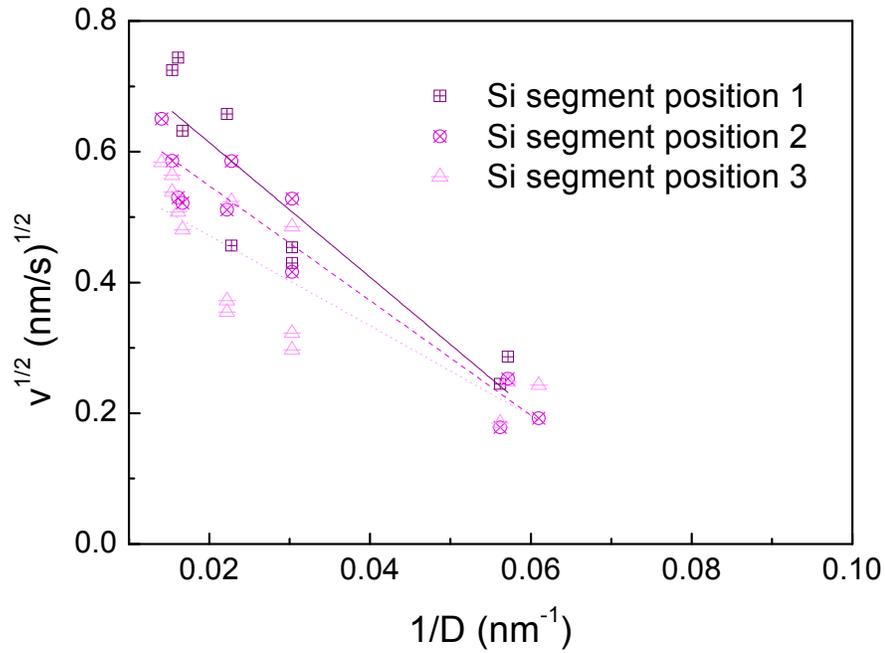

**Figure S4. Plot of the square root of the Si growth rate as a function of inverse diameter.** This data is obtained from wires grown with a 2 µm wire-to-wire spacing, such that the synergetic effect can be neglected. This plot illustrates the good agreement of Si nanowire growth with the Gibbs-Thomson effect for different positions along the superlattice GaP-Si nanowire.[31]

|  | **Critical diameter (nm)** |
| --- | --- |
| **Segment at position 1** | 12.6±3.0 |
| **Segment at position 2** | 12.1±1.3 |
| **Segment at position 3** | 11.3±2.7 |

**Table S1:** values of the critical diameter extracted from the fits in **fig. S5**. Below these values Si nanowires will not grow.

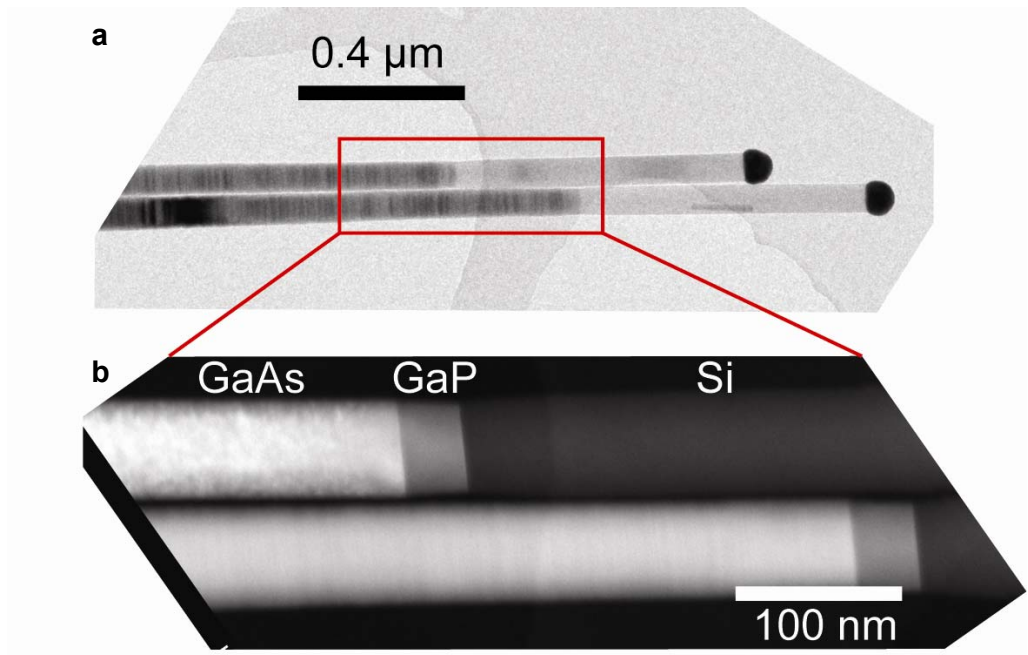

**Figure S5. Structural characterization of hybrid Si/GaAs nanowires. a.** Bright field TEM image and **b.** HAADF image of hybrid Si/GaAs nanowires. From these two images we see a long GaAs segment with a mixture of wurtzite and zincblende phases.